\documentclass[aps,twocolumn,prb,amsmath,amssymb,floatfix,superscriptaddress]{revtex4-2} 

\usepackage[utf8]{inputenc}
\usepackage{graphicx}
\usepackage{bm}
\usepackage{color}
\usepackage[colorlinks=true,citecolor=blue]{hyperref}
\usepackage{graphicx}
\usepackage{amssymb}

\usepackage{physics} 
\usepackage{comment}  
\makeatletter
\AtBeginDocument{\let\LS@rot\@undefined}
\makeatother

\begin{document}

\title{Odd-frequency  pairing due to  Majorana and    trivial Andreev bound states}

\author{Eslam Ahmed}
\affiliation{Department of Applied Physics, Nagoya University, Nagoya 464--8603, Japan}
\author{Shun Tamura}
\affiliation{Department of Physics, Nagoya University, Nagoya 464--8602, Japan}

\author{Yukio Tanaka}
\affiliation{Department of Applied Physics, Nagoya University, Nagoya 464--8603, Japan}
\affiliation{Research Center for Crystalline Materials Engineering, Nagoya University, Nagoya 464-8603, Japan}
\author{Jorge Cayao}
\affiliation{Department of Physics and Astronomy, Uppsala University, Box 516, S-751 20 Uppsala, Sweden}

\date{\today}

\begin{abstract}
 Majorana and trivial Andreev bound states are predicted to appear in superconductor-semiconductor hybrid systems, but their identification is still a challenging task. Here, we consider superconducting junctions with Rashba spin-orbit coupling and explore the signatures of Majorana and trivial Andreev bound states in the emergent superconducting correlations when the systems are subjected to an external Zeeman field. We first show that robust zero-energy Andreev bound states naturally appear due to confinement and helicity when the normal sector of the junction becomes helical. These Andreev states can evolve into Majorana states, developing alike oscillations around zero energy as a function of Zeeman field. Unlike Majorana states located at both ends, helical Andreev states are located at the interface. We then demonstrate that the emergent superconducting correlations are locally composed of even-frequency spin-singlet even-parity and odd-frequency spin-triplet even-parity pair amplitudes, which coexist due to the interplay of spin-orbit coupling, Zeeman field, and spatial translation invariance breaking. In the helical regime, trivial Andreev states enhance odd-frequency spin-triplet pairing, which decays in the superconductor and has a homogeneous long-range profile in the normal region. At zero frequency, however, odd-frequency spin-triplet pairing vanishes in the helical regime. In the topological phase, Majorana states enhance odd-frequency spin-triplet pairing, producing a long-range homogeneous leakage into the normal region. Interestingly, we discover that when Majorana states are truly zero-energy modes,  odd-frequency pairing develops a divergent low-frequency profile, which we interpret as the unambiguous self-conjugated Majorana signature. Our results help understand Majorana and trivial Andreev states from a superconducting correlations perspective in Majorana devices.
\end{abstract}
\maketitle

\section{Introduction}
Superconductor-semiconductor systems have attracted enormous attention in the past ten years largely due to their potential for realizing topological superconductivity \cite{sato2017topological,Aguadoreview17,lutchyn2018majorana,prada2019andreev,flensberg2021engineered,Marra_2022,tanaka2024theory}. This topological phase emerges when an external Zeeman field is applied and is characterized by the formation of self-conjugate quasiparticles known as Majorana bound states (MBSs) \cite{Aguadoreview17,tanaka2024theory}. MBSs appear at the ends of the system and acquire zero energy when the system is longer than their localization length \cite{sato2016majorana,Aguadoreview17,tanaka2024theory}. Since the first  experimental report in 2012, the zero energy Majorana nature has been mostly explored via conductance \cite{zhang2019next}, with interesting reports but without conclusive Majorana evidence so far \cite{prada2019andreev}. The main obstacle has been shown to be that superconductor-semiconductor systems naturally host topologically trivial zero-energy Andreev bound states (ABSs) \cite{Bagrets:PRL12,Pikulin2012A,PhysRevB.86.100503,PhysRevB.91.024514,PhysRevB.86.180503,PhysRevB.98.245407,PhysRevLett.123.117001,DasSarma2021Disorder,PhysRevB.105.144509,PhysRevB.104.L020501,PhysRevB.104.134507,marra2022majorana,PhysRevB.107.184509,baldo2023zero,PhysRevB.107.184519,PhysRevB.110.165404,PhysRevB.110.085414}, with properties similar to those of MBSs \cite{TK95,KT96,Kashiwaya_RPP,Proximityp,PhysRevLett.103.237001,PhysRevB.82.180516}. It is therefore imperative to explore other   properties to distinguish MBSs from trivial ABSs.

A less explored  property of MBSs is their self-conjugation, which is tied to their intrinsic  spatial nonlocality and charge neutrality \cite{Aguadoreview17,tanaka2024theory}. Interestingly, this self-conjugation has been shown to give rise to a superconducting correlation (or pair amplitude) that exhibits a divergent \emph{odd} frequency dependence \cite{golubov2009odd,tanaka2012symmetry,tanaka2018surface,cayao2019odd,mizushima2018multifaceted,tanaka2021theory,tanaka2024theory}, thereby signaling that the superconducting state with MBSs is a dynamical phase. This occurs because the intrinsic self-conjugation of MBSs gives rise to a particle-particle propagator that is the same as the particle-hole propagator and equal to $1/\omega$, where $\omega$ is the frequency.  Odd-frequency superconducting pairing has been explored in trivial \cite{odd3b,Eschrig2007,
RevModPhys.91.045005,triola2020role} and topological superconductors 
\cite{golubov2009odd,tanaka2012symmetry,tanaka2018surface,cayao2019odd,mizushima2018multifaceted,tanaka2021theory,tanaka2024theory}  and the divergent odd-frequency pairing has been shown to appear in the topological phase in the presence of MBSs, see also Refs.\,\cite{odd1,odd3,PhysRevB.87.104513,PhysRevB.92.121404,PhysRevB.95.184506,PhysRevB.96.155426,thanos2019,Takagi18,PhysRevB.101.214507,PhysRevB.101.094506,PhysRevB.106.L100502,PhysRevB.110.125408}.   Given that trivial ABSs mimic several properties of MBSs \cite{prada2019andreev}, and odd-frequency pairing stems from the self-conjugation of MBSs \cite{cayao2019odd}, it is natural to wonder if odd-frequency pairing offers a way to distinguish MBSs from trivial ABSs.  However, despite the efforts, it is still unknown what the behavior of odd-frequency pairing is like in systems where MBSs and trivial ABSs coexist, especially in Majorana devices based on superconductor-semiconductor hybrids. 

\begin{figure}[!t]
    \centering
    \includegraphics[width=\columnwidth]{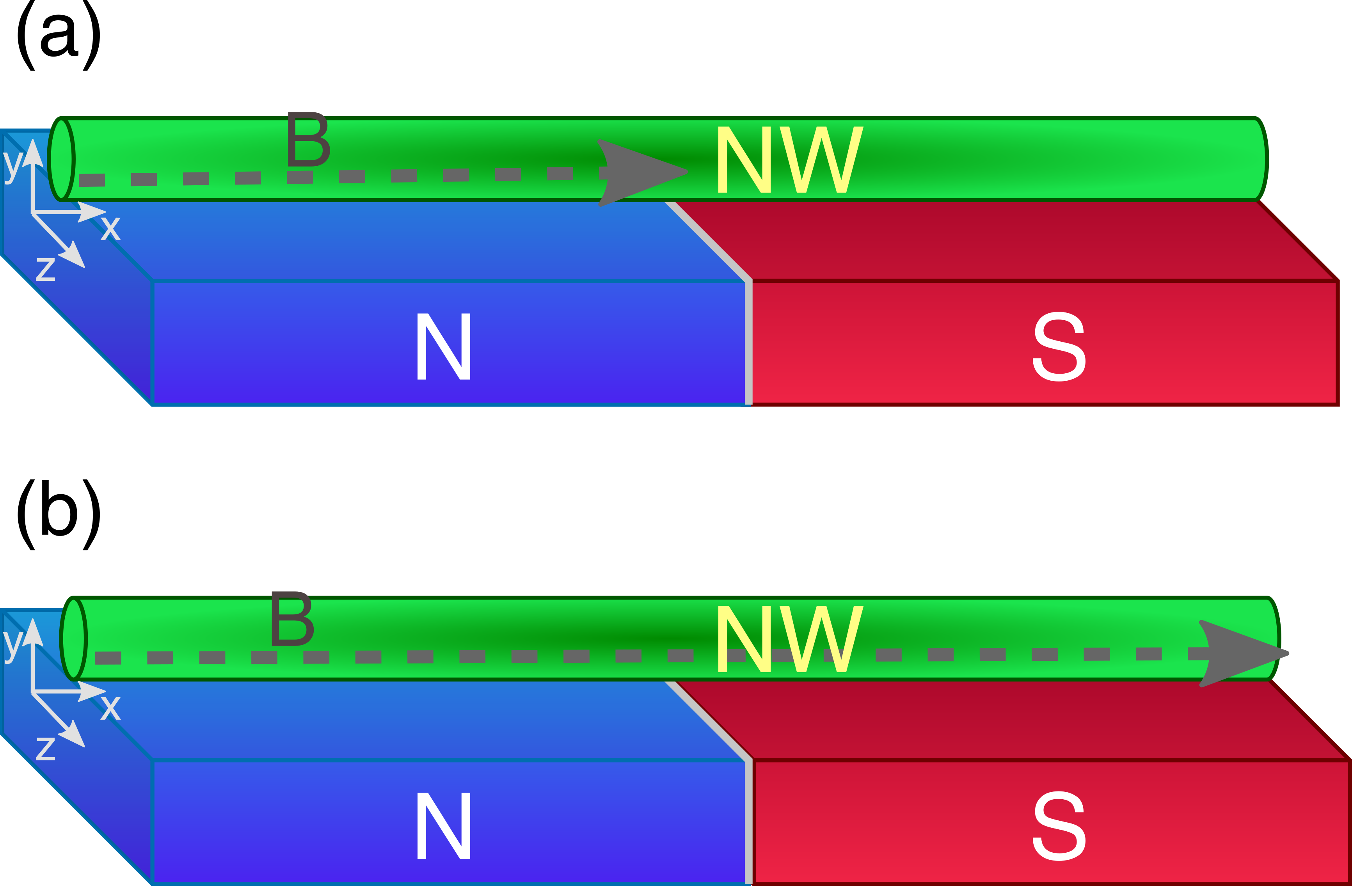}
    \caption{Schemes of the studied systems: (a) A semiconductor nanowire (NW, green) is placed on top of a superconductor (S, red) and a normal metal (N, blue). The N region has a finite Zeeman field $B$, indicated by a dashed horizontal arrow.    (b) The same junction as in (a) but with a Zeeman field applied over the entire NW.  In both cases, a very thin insulating layer (I, grey) is placed in the middle of the junction to avoid the leakage of Cooper pairs into the normal region and to ensure a homogeneous order parameter inside the superconducting region }
    \label{fig:junction}
\end{figure}

In this work, we consider normal-superconductor (NS) junctions with Rashba spin-orbit coupling (SOC)  and study the emergence of superconducting correlations in the presence of trivial ABSs and MBSs when the systems are subjected to an external Zeeman field. In particular, we consider two configurations of the Zeeman field in the NS junctions, which include first having a Zeeman field in N and then a Zeeman field over the entire N and S regions, see Fig.\,\ref{fig:junction}. We first demonstrate that  NS junctions in both configurations of Zeeman fields are able to host topologically trivial zero-energy ABSs due to the interplay of confinement and helicity in N. The helical phase appears well below the topological phase transition and the trivial ABSs emerge strongly located at the NS interface, with an exponentially decaying wavefunction in S and with regular spatial oscillations  in N. When the Zeeman field is over the entire system, we find that the trivial ABSs in the helical phase smoothly evolve towards MBSs in the topological phase, with both exhibiting a very similar oscillatory dependence on the Zeeman field for short S. We show that these Zeeman dependent oscillations only reduce for the MBSs when taking longer superconducting regions, which reflects the spatial nonlocality of MBSs located at both ends of S, while the trivial ABSs only appear at the NS interface.

We then show that, under general conditions,  locally even-frequency spin-singlet and odd-frequency spin-triplet pair amplitudes coexist due to the interplay between SOC, Zeeman field, and spatial translation invariance breaking at the NS interface. In the helical phase, the odd-frequency spin-triplet pair amplitudes develop large values at the NS interface, where a trivial ABS appears, while it acquires homogeneous oscillations in N and an exponentially decaying profile into the bulk of S. The even-frequency spin-singlet pair amplitude in the helical phase leaks into N, while exhibiting an overall finite value in S due to the spin-singlet $s$-wave nature of the parent superconductor.  In the topological phase, the odd-frequency spin-triplet component takes values larger than those of the spin-singlet even-frequency part,
with a profile having homogeneous oscillations in N and an oscillatory exponential decay  from both ends of S due to MBSs.  The finite value of the spin-singlet and spin-triplet pair correlations in N means that they determine the superconducting proximity effect, which involves a large amount of odd-frequency spin-triplet pairing deep in N at low frequencies in the helical and topological phases. Similarly, the finite values of odd-frequency spin-triplet pair amplitudes in S indicate their contribution to the inverse proximity effect, which is most prominent in the topological phase due to the presence of MBSs. 

When inspecting the frequency dependence, we find that the odd-frequency spin-triplet pair amplitudes in the helical phase are linear in Matsubara frequency, hence vanishing at zero frequency but with large values at the energy of the trivial ABSs. In the topological phase, we obtain that the odd-frequency spin-triplet pair amplitude develops a distinct behavior depending on whether the length of the S sectors $L_{\rm S}$ is shorter or longer than the  localization length of MBSs $\ell_{\rm M}$. For short S regions $L_{\rm S}\leq2\ell_{\rm M}$, the odd-frequency pair amplitude acquires a linear dependence in frequency as $\sim\omega_{n}$ and therefore vanishes at zero Matsubara frequency, in the same way as for trivial zero-energy ABSs. In the case of  long S regions $L_{\rm S}\gg2\ell_{\rm M}$, we show that the odd-frequency pair amplitude diverges near zero frequencies as $\sim1/\omega_{n}$, which we interpret as a signature of MBSs becoming truly zero modes.  Since zero-energy MBSs for $L_{\rm S}\gg2\ell_{\rm M}$ are really nonlocal in space, with their wavefunctions not overlapping, the odd-frequency spin-triplet pair amplitudes having a $\sim1/\omega_{n}$ dependence reveal the inherent charge neutrality and self-conjugation of MBSs.

We   also assess  the effect of tilting the Zeeman field on the helical and topological phases, which can be naturally present under realistic conditions.  We obtain that trivial ABSs are ubiquitously present for all orientations of the Zeeman field except when it is parallel to the Rashba SOC, where the helical phase is suppressed due to the absence of a helical gap.  In the topological phase, the bulk gap closes exactly when the Zeeman field component parallel to the SOC, implying that there is a critical field below which MBSs with topological protection appear.
In   case of the superconducting correlations, the triplet components are sensitive to the orientation of the Zeeman field, showing an oscillatory behavior as a function of the angle when the Zeeman field is rotating around the Rashba SO axis. 
 
The remainder of this work is organized as follows. In Section~\ref{sec2} we discuss the  NS junction models with the two Zeeman field configurations and also outline the method to obtain the superconducting correlations using Green's functions.
In  Section~\ref{section3} we show the formation of trivial ABSs and analyze the emergent pair correlations in an NS junction with a Zeeman field in N, while in Section~\ref{section4} we do the same but with a Zeeman field in both N and S.
  Finally, in Section~\ref{section5}, we present our conclusions.  

\section{Model and methods}
\label{sec2}
We consider  a one-dimensional (1D) semiconducting nanowire (NW) with  intrinsic Rashba SOC, where part of it is  placed in contact with a normal metal (N) and the other part in contact with a conventional $s$-wave superconductor (S), see Fig.\,\ref{fig:junction}. As a result,  an $s$-wave superconducting pairing potential is induced inside the NW by   proximity effect, which originates an NS junction.  Moreover, we assume that a Zeeman field is induced, e.g., due to an external Zeeman field or a ferromagnet (F), either   in only the N or in both N and S, as depicted in Fig.\,\ref{fig:junction}.   We model our setup with the following Bogoliubov-de Gennes Hamiltonian
    \begin{equation}\label{eq: hamiltonian}
        H = H_{\rm NW} + H_{\rm B} + H_{\Delta}\,,
    \end{equation}
    where
\begin{equation}
\begin{split}
H_{\rm NW} &= \sum_{j=1}^N\psi_j^\dagger\left(\left(2t-\mu(j)\right)\tau_z\sigma_0 \right)\psi_j   \\
 &+ \sum_{j=1}^{N-1}\psi_j^\dagger\left(-t\tau_z\sigma_0 - i\frac{\alpha}{2a}\tau_0\sigma_z\right)\psi_{j+1}+h.c.\,,\\
         H_{\rm B} &= \sum_{j=1}^N\psi_j^\dagger\left( - B(j)\tau_z\sigma_x \right)\psi_j\,,\\
        H_{\Delta} &= \sum_{j=L_{\rm N}/a+1}^N\psi_j^\dagger\left(  \Delta\tau_y\sigma_y\right)\psi_j\,,
\end{split}
\end{equation}
where $\psi_j=(c_{j\uparrow},c_{j\downarrow},c^\dagger_{j\uparrow},c^\dagger_{j\downarrow})^\mathrm{T}$ is the Nambu spinor at site $j$, $c_{j\sigma}$ ($c_{j\sigma}^{\dagger}$) destroys (creates) a fermionic state at site $j$ with spin $\sigma$, $\tau_{i}\left(\sigma_{i}\right)$ is the $i$-th Pauli matrix in Nambu(spin) space, $t$ is the nearest-neighbor hopping amplitude, $\alpha$ is the Rashba SOC strength, $\Delta$ is the $s$-wave pairing order parameter, $B(j)$ is the Zeeman coupling at site $j$,  and $\mu(j)$ is the chemical potential at site $j$. The length of the system is given by $L=Na$, where $N$ represents the number of sites and $a$ the lattice spacing. We restrict the pair potential to only the right  region of the NW, thus realizing an NS junction, see Fig.\ref{fig:junction}. We set the length of the S (N) region to $L_{\rm S} (L_{\rm N})$ such that $L=L_{\rm S}+L_{\rm N}$. We allow the chemical potential to take different values in each region such that
\begin{equation}
    \mu(j)=\begin{cases}
        \mu_{\rm N}\,, & 1\leq ja\leq L_{\rm N}\,,\\
        \mu_{\rm S}\,, & L_{\rm N}< ja\leq L\,.
    \end{cases}
\end{equation}
Since we are interested in exploring the emergent superconducting correlations in NS junctions with MBSs and trivial zero-energy ABSs, we consider two different spatial profiles of $B(j)$  where they are very likely to appear. First, we restrict the Zeeman field to the N region only, such that $B(j)=B\theta(L_{\rm N}-ja)\theta(ja)$ [Fig.\ref{fig:junction}(a)], which could happen e.g., when placing a F in partial contact with N: this scenario is capable of only hosting trivial ABSs but does not host MBSs. Second,  we consider an NS junction with  a Zeeman field constant all over the N and S regions, such that $B(j)=B$ [Fig.\ref{fig:junction}(b)]; this could be due to e.g.\, a F or an external Zeeman field. This second system [Fig.\ref{fig:junction}(b)] is able to host both trivial zero-energy ABSs and also MBSs, which appear in distinct phases driven by the Zeeman field.   For  $B>B_{\rm h}\equiv \mu_{\rm N}$, the N region hosts counter-propagating states (two Fermi points) and thus becomes helical with the possibility to host robust trivial zero-energy ABSs \cite{PhysRevB.91.024514,JorgeEPs,PhysRevB.104.L020501,baldo2023zero}. For $B>B_{\rm c}\equiv \sqrt{\mu_{\rm S}^{2}+\Delta^{2}}$ the S region becomes topological and hosts MBSs \cite{tanaka2024theory}. The helical phase also appears in the first system [Fig.\ref{fig:junction}(a)], which justifies its consideration. The two systems discussed here are expected to form under realistic conditions in Majorana devices \cite{prada2019andreev,frolov2019quest,flensberg2021engineered}, where we aim to identify the nature of superconducting correlations in the presence of trivial zero-energy ABSs and MBSs.

\subsection{Superconducting correlations}
\label{subsection2a}
To describe the emergent superconducting correlations in the NS junctions discussed before, we calculate the anomalous retarded (\textit{r}) Green's function  \cite{mahan2013many,zagoskin}, 
\begin{equation}
\label{Fret}
\mathcal{F}_{jj',\sigma\sigma'}^{r}(t,t')=-i\theta(t-t')\ev{\acomm{c_{j\sigma}(t)}{ c_{j'\sigma'}(t')}}
\end{equation}
where $c_{j\sigma}(t)$ destroys a 
fermionic state at   site $j$, spin $\sigma$,   at time $t$. For completeness, we note that in what follows, we will refer to $\mathcal{F}_{jj',\sigma\sigma'}^{r}(t,t')$ as   pair amplitude, pair correlation, or superconducting pair correlation. As shown by Eq.\,(\ref{Fret}), the pair amplitude in general involves pairing of fermionic states at distinct times and involving all the quantum numbers of the paired operators. The advanced (\textit{a}) pair amplitude can also be found as $\mathcal{F}_{jj',\sigma\sigma'}^{a}(t,t')=i\theta(t'-t)\ev{\acomm{c_{j\sigma}(t)}{ c_{j'\sigma'}(t')}}$. Since we are dealing with a stationary case, we can set $t'=0$ and Fourier transform into frequency space as
\begin{equation}
\label{Fw}
 F^{r(a)}_{jj',\sigma\sigma'}(\omega)=\int\mathcal{F}_{jj',\sigma\sigma'}^{r(a)}(t,0)\,e^{i[\omega +(-) i\delta]t}\,dt
\end{equation}
for the retarded (advanced) pair amplitudes, with $\delta$ being an infinitesimally small positive number. At this point, we recall that the pair amplitude describes the pairing of two fermionic states, which implies that it is antisymmetric under the total exchange of quantum numbers including frequency \cite{RevModPhys.77.1321,tanaka2012symmetry,
Balatsky2017,cayao2019odd,triola2020role,tanaka2021theory}. For the retarded and advanced pair amplitudes, this antisymmetry condition is given by $F^{r}_{jj',\sigma\sigma'}(\omega)=-F^{a}_{j'j,\sigma'\sigma}(-\omega)$, where we passed to the advanced branch when   $\omega\rightarrow-\omega$. While the pair amplitude is antisymmetric under a total exchange of quantum numbers, it can be symmetric (even) or antisymmetric (odd) under the individual exchange of   e.g.\, spins, lattice sites, or frequency. Thus, under the exchange of spins, the pair amplitude can be \emph{singlet} or \emph{triplet}, while   under the exchange of lattice sites and frequency  the pair amplitude can be \emph{even} or \emph{odd}. Under general circumstances, there are four pair symmetry classes allowed by the antisymmetry condition but they can be reduced by further assuming some specifics about the systems with MBSs and ABSs we consider. Since there is theoretical evidence that MBSs and ABSs appear localized  at boundaries \cite{odd3,tanaka2012symmetry}, it is reasonable to address the pair amplitudes locally in space, namely, at $j=j'$; thus, the exchange of site indices only gives an even parity symmetry. Taking this  into account,  only two pair symmetries are allowed: i) even-frequency spin-singlet even-parity (ESE), and ii) odd-frequency spin-triplet even-parity (OTE) \cite{Berezinskii74,Coleman93,
tanaka2012symmetry,Balatsky2017}. 
In practice, we calculate $F^{r(a)}_{jj',\sigma\sigma'}(\omega)$ from the off-diagonals of the  Nambu Green's function in frequency domain as
\begin{equation}
\label{GG}
    \mathcal{G}^{r(a)}_{jj',\sigma\sigma'}(\omega) = \left[\omega+(-) i\delta-H\right]^{-1}_{jj',\sigma\sigma'} \,,
\end{equation}
where $H$ is given by Eq.\,(\ref{eq: hamiltonian}) and describes the respective NS junctions under consideration. The structure of $\mathcal{G}^{r(a)}_{jj',\sigma\sigma'}(\omega)$ in Nambu space is given by
\begin{equation}
    \mathcal{G}^{r(a)}_{jj',\sigma\sigma'}(\omega)= \begin{pmatrix}
        G^{r(a)}_{jj',\sigma\sigma'}(\omega) & F^{r(a)}_{jj',\sigma\sigma'}(\omega)\\
        \bar{F}^{r(a)}_{jj',\sigma\sigma'}(\omega) & \bar{G}^{r(a)}_{jj',\sigma\sigma'}(\omega)
    \end{pmatrix}\,,
\end{equation}
where the diagonal elements, $G^{r(a)}_{jj',\sigma\sigma'}(\omega)$ and  $\bar{G}^{r(a)}_{jj',\sigma\sigma'}(\omega)$, represent the normal electron-electron and hole-hole Green's functions, respectively;   the off-diagonal elements, $F^{r(a)}_{jj',\sigma\sigma'}(\omega)$ and $\bar{F}^{r(a)}_{jj',\sigma\sigma'}(\omega)$, are the anomalous electron-hole and hole-electron Green's functions, respectively, that describe the pair amplitudes. Given that the systems under consideration include SOC and a Zeeman field, we expect to find spin-singlet pairing as well as mixed spin-triplet and equal spin-triplet pairings. 

Upon symmetrization over frequency, we calculate the ESE and the   OTE pair amplitudes as \cite{cayao2019odd,RevModPhys.91.045005}
\begin{equation}
\label{EOpairamplitudes}
\begin{split}
   F^{{\rm E}}_{\sigma\sigma',j}(\omega) &= \frac{F^r_{jj,\sigma\sigma'}(\omega)+F^a_{jj,\sigma\sigma'}(-\omega)}{2}\,,\\ 
    F^{{\rm O}}_{\sigma\sigma',j}(\omega) &= \frac{F^r_{jj,\sigma\sigma'}(\omega)-F^a_{jj,\sigma\sigma'}(-\omega)}{2}\,,\\   
\end{split}
\end{equation}
where $\sigma,\sigma'=\{\uparrow,\downarrow\}$. Thus, the   pair amplitude $F^{{\rm E}}_{\uparrow\downarrow,j}(\omega)$ represents the ESE pair symmetry class, while $F^{{\rm O}}_{\uparrow\downarrow,j}(\omega)$
and $ F^{{\rm O}}_{\sigma\sigma,j}(\omega)$ are the mixed and equal spin-triplet OTE classes, respectively. It is also convenient to characterize the pair amplitude from the off-diagonal parts of the Matsubara Green's function in Nambu space, obtained from $\mathcal{G}^r_{jj',\sigma\sigma'}(\omega)$ given in Eq.\,(\ref{GG}) by replacing $\omega+i\delta\rightarrow i\omega_{n}$, with $\omega_{n}$ being the Matsubara frequency. Within Matsubara representation, the antisymmetry condition discussed below Eq.\,(\ref{Fw}) reads $\mathsf{F}_{jj'}^{\sigma\sigma'}(i\omega_{n})=-\mathsf{F}_{j'j}^{\sigma'\sigma}(-i\omega_{n})$, where $\mathsf{F}_{jj'}^{\sigma\sigma'}(i\omega_{n})$ denotes the  pair amplitude in Matsubara representation. Then, the even- and odd-frequency pair amplitudes at $j=j'$ in Matsubara representation can be obtained as
\begin{equation}
\label{Fmatsub}
\begin{split}
\mathsf{F}^{\rm E}_{\sigma\sigma',j}(i\omega_{n})&=\frac{\mathsf{F}_{jj}^{\sigma\sigma'}(i\omega_{n})+\mathsf{F}_{jj}^{\sigma\sigma'}(-i\omega_{n})}{2}\,,\\
\mathsf{F}^{\rm O}_{\sigma\sigma',j}(i\omega_{n})&=\frac{\mathsf{F}_{jj}^{\sigma\sigma'}(i\omega_{n})-\mathsf{F}_{jj}^{\sigma\sigma'}(-i\omega_{n})}{2}\,.
\end{split}
\end{equation}
Analogously to Eqs.\,(\ref{EOpairamplitudes}), $\mathsf{F}^{\rm E}_{\uparrow\downarrow,j}(i\omega_{n})$ is the ESE pair amplitude, while $\mathsf{F}^{\rm O}_{\uparrow\downarrow,j}(i\omega_{n})$ and $\mathsf{F}^{\rm O}_{\sigma\sigma,j}(i\omega_{n})$ are the mixed and equal spin-triplet OTE classes.   Thus, under general circumstances, the anomalous Green's function (either retarded or Matsubara) is a matrix in spin space, and its elements being the pair amplitudes follow Eqs.\,(\ref{EOpairamplitudes}) for the retarded functions or Eqs.\,(\ref{Fmatsub}) for the Matsubara counterparts \cite{Takagi2022}.  In both cases, it is also common to write the anomalous Green's function as
\begin{equation}
\mathbf{F}=(d_{s}\sigma_{0}+\mathbf{d}\cdot\boldsymbol{\sigma})(i\sigma_{y})
\end{equation}
where $d_{s,x,y,z}$ are given by either Eqs.\,(\ref{EOpairamplitudes})  or Eqs.\,(\ref{Fmatsub}) depending on whether we are dealing with the retarded or Matsubara pair amplitudes. For the retarded pair amplitudes, we have $d_{s}=F^{\rm E}_{\uparrow\downarrow,j}$, 
$\mathbf{d}=(d_{x},d_{y},d_{z})$, with 
$d_{x}=(F^{\rm O}_{\downarrow\downarrow,j}-F^{\rm O}_{\uparrow\uparrow,j})/2$, $d_{y}=i(F^{\rm O}_{\downarrow\downarrow,j}+F^{\rm O}_{\uparrow\uparrow,j})/2$, 
$d_{z}=F^{\rm O}_{\uparrow\downarrow,j}$. Similar expressions hold in the Matsubara representation. 

 In what follows, we numerically calculate the emergent pair amplitudes given by Eqs.\,(\ref{EOpairamplitudes}) for   NS junctions with MBSs and trivial ABSs described by Eq.\,(\ref{eq: hamiltonian}). Whenever necessary we will also make use of Eqs.\,(\ref{Fmatsub}) to inspect the emergent superconducting correlations.  We consider realistic parameters, which include $\alpha=20$\,meVnm and  $\Delta=0.25$\,meV, according to the experimental values reported for   InSb and InAs nanowires and Nb and Al superconductors \cite{lutchyn2018majorana}. Moreover, we also take $a=10$\,nm, $t=25$\,meV, $\mu_{\rm S}=0.5$\,meV. In Sec.~\ref{section3}, we set $\mu_{\rm N}=0.5$\,meV while, in Sec.~\ref{section4}, we take $\mu_{\rm N}=0.1$\,meV unless otherwise specified.

\begin{figure*}[!th]
    \centering
    \includegraphics[width=\textwidth]{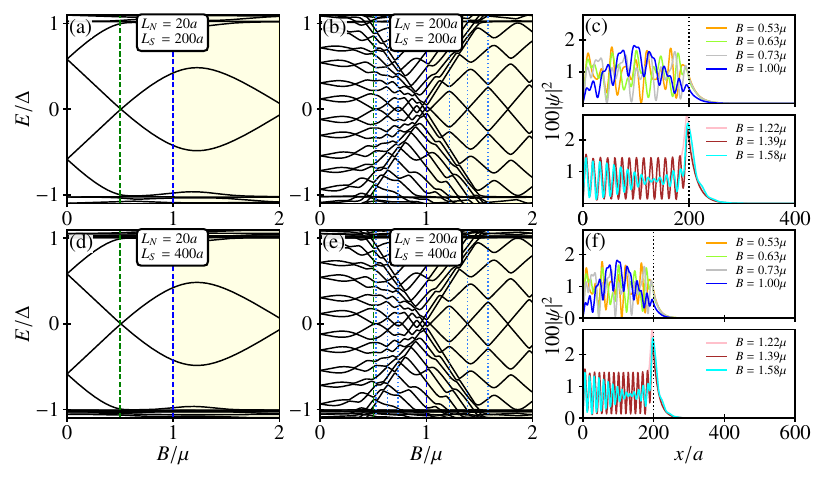}
    \caption{(a,b) Low energy spectrum for an NS junction with the Zeeman field  $B$ in N as a function of  $B$ for $L_{\rm N}=20a$ and $L_{\rm N}=200a$, both at $L_{\rm S}=200a$.         (c) Wavefunction probability density of the state closest to zero energy as a function of the spatial coordinate at distinct values of the Zeeman field for $L_\mathrm{N,S}=200a$; curves in the top panel correspond to $B\leq\mu$, whereas those in the bottom panel correspond to $B>\mu$, with $\mu=\mu_{\rm N,S}$. (d,e), and (f) Same as in (a,b) and (c), respectively,  but for $L_{\rm S}=400a$. The green and blue dashed vertical lines in (a,b,d,e) mark $B=\Delta$ and $B=\mu$, respectively, with $B\geq\mu$ in yellow color;     the dotted curves in (b,e) indicate the values of $B$ at which the wavefunction probability densities are plotted in (c,f). 
Parameters:  $a=10$\,nm, $\mu_{\rm N,S} \equiv \mu = 0.5$\,meV, $\Delta = 0.25$\,meV, $t=25$\,meV, $\alpha = 20$\, meV nm. }
    \label{Fig2}
\end{figure*}

\section{Hybrid system with trivial zero-energy ABSs but without MBSs}
\label{section3}
We begin by investigating the NS junction with a Zeeman field only in N and hence hosting only trivial ABS, see Fig.\,\ref{fig:junction}(a). This particular setup prevents the condition necessary to form MBSs, which allows us to focus solely on trivial ABS. Throughout this section, we fix the chemical potential  to $\mu_{\rm N,S}\equiv\mu=0.5$\,meV. Thus, the N region becomes helical at $B_{h}\equiv\mu_{\rm N}$ \cite{PhysRevB.91.024514}. To gain a detailed understanding, we first analyze the formation of trivial ABSs and then we explore how they influence the  emergent superconducting correlations.

\subsection{Low-energy Spectrum}
The energy spectrum is numerically obtained by diagonalizing the BdG Hamiltonian given by Eq.\,(\ref{eq: hamiltonian}) for realistic parameters. In Fig.\,\ref{Fig2} we present the low-energy spectrum as a function of the Zeeman field $ B$ for short and long N regions with $L_{\rm N}=20a$ and $L_{\rm N}=200a$, respectively, in both cases for $L_{\rm S}=200a$ and $L_{\rm S}=400a$.  In Fig.\,\ref{Fig2}(c,f) we also present the wavefunction probability density of the  state closest to zero energy $|\psi(x)|^{2}=|\psi_E(x)|^2 + |\psi_{-E}(x)|^2$ as a function of the spatial coordinate $x$ at distinct values of the Zeeman field marked by dotted vertical lines in Fig.\,\ref{Fig2}(b,e). The wavefunction $\psi_E(x)$ corresponds to the eigenstate of Eq.\,\eqref{eq: hamiltonian} at the position $x=ja$ and energy $E$ and has the following structure $\psi_E(x=ja)=(u^\uparrow_j,u^\downarrow_j,v^\uparrow_j,v^\downarrow_j)^T$, where $u^\sigma_j$ and $v^\sigma_j$ are the electron and hole components, respectively.

In the case of short N regions [Fig.\,\ref{Fig2}(a,d)], the junction hosts a pair of in-gap energy levels as a result of confinement due to the finite length of N but independent of the superconducting length. These in-gap states split at finite Zeeman field $B$ and, as $B$ increases, the electron and hole energy components with opposite spin oscillate with the Zeeman field developing a first zero energy crossing  at $B=\Delta$, see vertical green dashed line.  As the Zeeman field further increases and traverses $B=\mu$, the in-gap spectrum does not suffer any influence other than the pair of in-gap levels continuing to oscillate within a well-defined energy gap that separates them from the quasicontinuum. It is worth noting that $B=\mu$ marks the transition into the helical  regime (yellow shaded region) where there are only two counter-propagating states with opposite spins formed in the N and S regions; however, the in-gap spectrum of NS junctions with short N regions is not influenced neither by the helical transition nor by longer S regions.

For long N segments  [Fig.\,\ref{Fig2}(b,e)] the situation is more interesting as several   energy levels appear within the gap which then develop a nontrivial behavior with the increase of the Zeeman field. All in-gap levels with both spin components oscillate as $B$ takes finite values and develop an intriguing behavior at low and high energies.  In the low-energy sector, the lowest energy levels reach zero energy after $B=\Delta$ in an oscillatory fashion which is however not periodic until $B=\mu$. At high energies, there appears a pronounced set of levels after $B=\Delta$  that moves towards zero energy linearly with $B$, forming a sort of gap closing and reopening feature at $B=\mu$ as a result of the emergence of the helical phase but without any relation to topology \cite{PhysRevB.91.024514}. In the helical phase, for $B>\mu$, one spin sector gets removed and the low-energy sector represents a spin-polarized spectrum with zero-energy crossings. The behavior of the low-energy spectrum in Fig.\,\ref{Fig2} reveals that  trivial zero-energy ABSs indeed form without any relation to topology or Majorana physics in NS junctions. By comparing  Fig.\,\ref{Fig2}(b) and Fig.\,\ref{Fig2}(e), we also uncover that the in-gap energy spectrum does not depend on the length of the S region $L_{\rm S}$, showing that the localization of the wavefunctions is very likely to be at the interface.

To support the wavefunction localization, in Fig.\,\ref{Fig2}(c,f) we present the wavefunction probability densities $|\Psi|^{2}$ of the levels closest to zero energy corresponding to Fig.\,\ref{Fig2}(b,e). For Zeeman fields below the helical transition $B<\mu$, the wavefunction probability density in N exhibits an irregular oscillatory profile with a tiny beating feature due to the combined effect of SOC, Zeeman field, and chemical potential; here four Fermi momenta define the oscillatory pattern of $|\Psi|^{2}$. In the S region, the wavefunction probability decays in an exponential fashion with very small oscillations defined by SOC and chemical potential since there is no Zeeman field in S. In the helical phase, $B>\mu$, the wavefunction probabilities develop  regular oscillations with equal periodicity in N determined by the size of $B$. The number of oscillations in the N region is independent of $L_{\rm S}$, signaling that its origin stems from the normal state. Moreover, here we also find   a pronounced beating feature in $|\Psi|^{2}$,  appearing at $B$ fields when the pair of lowest energy levels  is closer in energy to the first excited state, see  $B=1.22\mu$ and $B=1.58\mu$.  At the NS interface in the helical phase,  $|\Psi|^{2}$ acquires a large value which shows the localization of a trivial ABS. This bound state  benefits from the competition between   superconductivity and the Zeeman field in N.  When entering into S, such a bound state decays exponentially with very small oscillations which are, however, difficult to identify by the naked eye.

\begin{figure*}[!th]
    \centering
    \includegraphics[width=\textwidth]{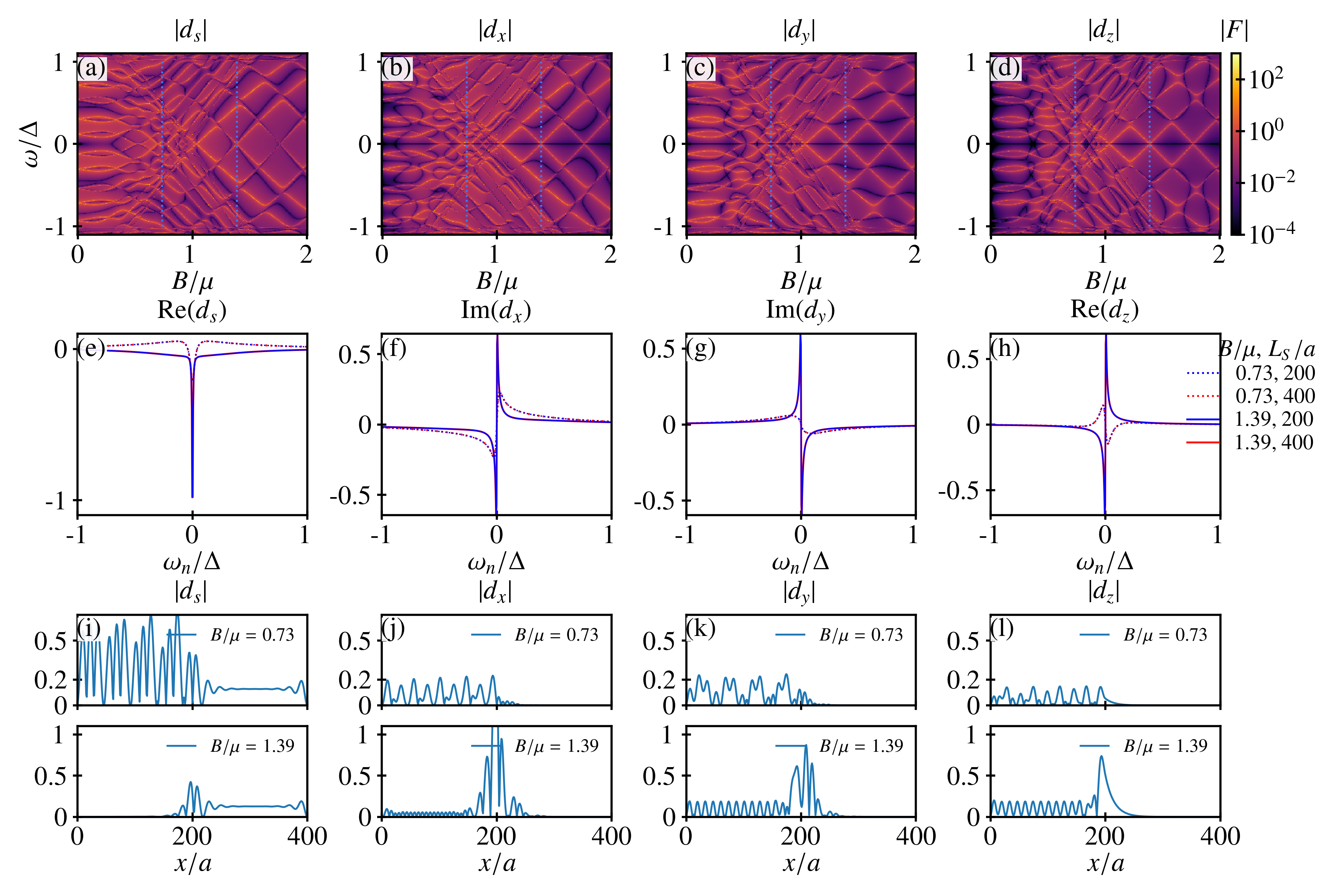}
    \caption{(a-d) Absolute value of the retarded  pair amplitudes  for an NS junction with the Zeeman field B in N as a function of frequency $\omega$ and   $B$ near the junction interface in N.  (e-h) Pair amplitudes from (a-d) as a function of Matsubara frequency $\omega_{n}$ at $x=192a$ for two distinct values of $B$ and $L_{\rm S}$; the curves below (above) the helical transition are  depicted in dashed (solid), while for a short (long) S region the curves are shown in blue (red). (i-l) Spatial profile of the pair amplitudes in (a-d) at $\omega\approx0$ and for   $B$ below and above the helical transition. The Zeeman field $B$ in the top panel of (i-l) is fixed at $B=0.73\mu$ which is below the helical transition, while the Zeeman field in the bottom panel is chosen above the helical transition $B=1.39\mu$. Parameters: $x=192a$, $L_{\rm N}=200a$, while $L_{\rm S}=200a$ unless otherwise specified. The rest of the parameters are the same as in Fig.\,\ref{Fig2}.}
    \label{Fig3}
\end{figure*}

\subsection{Emergent superconducting correlations}
Having established the regime with trivial zero-energy ABSs, we now devote our attention to explore how they impact the emergent superconducting correlations. For this purpose, we consider the regime with trivial zero-energy ABSs found in the previous subsection, which corresponds to a realistic NS junction with $L_{\rm N}=200a$, $L_{\rm S}=200a$. Then, to obtain the pair amplitudes, we follow the approach outlined in Subsection \ref{subsection2a}, taking into account   realistic parameters.  In Fig.\,\ref{Fig3}(a-d) we present the absolute value of the spin-singlet and spin-triplet retarded pair amplitudes as a function of frequency $\omega$ and Zeeman field $B$ near the interface of the junction in N and $j=j'$. Fig.\,\ref{Fig3}(e-h) shows the  pair amplitudes   as a function of Matsubara frequency $\omega_{n}$ at the same position of (a-d) for $B$ below and above the helical transition and for two distinct lengths of $L_{\rm S}$. Moreover, Fig.\,\ref{Fig3}(i-l) shows the absolute value of the spin-singlet and spin-triplet pair amplitudes as a function of space at $\omega\approx0$; therein we present two cases of Zeeman fields, below and above the helical transition $B=\mu$.

The first observation in Fig.\,\ref{Fig3}(a-d) is that the spin-singlet pair amplitude coexists at the interface with mixed spin-triplet and equal spin-triplet pair correlations. The singlet component corresponds to the ESE pair symmetry class, while the spin-triplet parts exhibit  OTE symmetry, see Subsection \ref{subsection2a}. Thus, the presence of OTE pair symmetry is consistent with the antisymmetry condition of superconducting correlations. While the ESE component stems from the spin-singlet parent superconductor, the OTE pairings are entirely induced due to the combined effect of spin-singlet superconductivity, SOC, and Zeeman field. All these interface  pair amplitudes exhibit high intensity values  that reveal the energy spectrum shown in Fig.\,\ref{Fig2}.  The ESE pair amplitude develops  roughly large values within the superconducting gap irrespective of $B$, while the OTE pair amplitudes become more pronounced as $B$ increases; it is interesting to see that at the interface the mixed spin-triplet OTE has an overall smaller size than the equal spin-triplet OTE parts.  Fig.\,\ref{Fig3}(a-d) also show that the OTE pair amplitudes exhibit reduced values as the frequency approaches zero. Further insights into the frequency profile are obtained from Fig.\,\ref{Fig3}(e-h) where we show the  dependence on Matsubara frequency $\omega_{n}$ of the singlet and triplet components for distinct values of $L_{\rm S}$ at fixed $B$; note that we show the pair amplitudes as a function of Matsubara frequency since the frequency dependence is simpler to identify. As expected, the ESE and OTE pair amplitudes are indeed   even and odd functions of frequency, with the OTE amplitudes developing large values near zero frequency but vanishing at $\omega_{n}=0$. This supports that the OTE pair amplitude has a linear dependence in frequency  ($\sim \omega_{n}$), with a rather large slope that makes it exhibit large values around zero frequency. The large slope arises because the chosen value of $B$ in the helical phase is at the zero-energy crossing; we have verified that, away from this zero-energy crossing, the slope is smaller and the linear frequency dependence of the OTE pair amplitudes is more visible. As the length of S increases, the OTE pair amplitudes do not change, a signature showing their spatial locality and their independence of the S size. The large OTE values at low frequencies can be understood to be   the result of the interplay between Zeeman field and SOC  with the spatial translation invariance breaking
promoting spin triplet $s$-wave Cooper pairs \cite{cayao2019odd}.  It is worth noting that the linear frequency dependence of the OTE pair amplitude was shown before to be common in topologically trivial systems \cite{PhysRevB.93.201402,PhysRevB.98.075425,PhysRevB.98.161408,Spectralbulk,cayao2019odd,triola2020role,PhysRevB.102.100507,PhysRevB.103.104505,PhysRevB.104.054519,PhysRevLett.129.247001,heinrich2024}. Thus, the regime with trivial zero-energy ABSs located at the NS interface exhibits an OTE pairing that is not divergent as it happens in systems with MBSs. The $\omega$-dependence of the OTE pairing signals the topologically trivial nature of the zero-energy ABSs found in Fig.\,\ref{Fig2}. 
 
Further understanding of the OTE pairing under the trivial zero-energy ABSs is obtained from  Fig.\,\ref{Fig3}(e-h), which shows the spatial dependence of the emergent pair amplitudes near zero frequency. Below the helical transition $B<\mu$, the ESE and OTE pair amplitudes acquire finite values in the N region, which decay from the interface  towards the bulk of S following an exponentially oscillatory decay; see top panels in Fig.\,\ref{Fig3}(e-h). The finite values of ESE and OTE in N signal the proximity effect, while the finite values of OTE pairing in S can be interpreted as the inverse proximity effect. Both pair amplitudes in N develop an irregular oscillatory profile with a beating pattern that  depends on the SOC, Zeeman field, and chemical potential; the origin of the irregular oscillations is the same as the one seen in the wavefunctions in Fig.\,\ref{Fig2}(c,f). In the S region, the ESE component exhibits a small decay from the interface but reaches a finite value in the bulk which is determined by the size of the pair potential since it is also of ESE symmetry. The OTE pairings exponentially decay from the interface with marked oscillations  that also depend on the Zeeman field, SOC, and chemical potential as in the N region. Deep in S, the OTE pair correlations become vanishingly small, although the OTE pairing always appears with an overall smaller size than the ESE component.

Above the helical transition $B>\mu$, the behavior of the pair amplitude has some similarities but also differences in comparison to the regime below the helical regime,  see bottom panels in Fig.\,\ref{Fig3}(e-h). There appear finite ESE and OTE pair amplitudes in N and S, supporting their contribution to the proximity effect as well as to  inverse proximity effect in the helical phase. For $B>\mu$ in the N region,   the OTE pair amplitudes develop   an  oscillatory homogeneous profile with larger values and a regular periodicity but without any beating feature, unlike the case for  $B<\mu$. The profile of the OTE pairing is akin to the behavior of the wavefunction in Fig.\,\ref{Fig2}(c,f). Interestingly, for strong Zeeman fields such as in the helical regime, the $x$-component of the $d$ vector is suppressed in the N region, while the $y$- and $z$-components are not, see the lower panels of Fig.\,\ref{Fig3}(j-l). This can be understood in terms of short- and long-range triplet correlations. Since the Zeeman field points in the $x$-direction, it tends to break pairs of the form $\ket{\rightarrow\leftarrow}+\ket{\leftarrow\rightarrow}$, which corresponds to the $x$-component of the $d$ vector. Here, $\ket{\rightarrow},\ket{\leftarrow}$ are the spin eigenstates of $\sigma_x$. Meanwhile, the $y$- and $z$-components of the $d$ vector represent long-range triplet correlations that can penetrate deeper into the N region, consistent with previous studies on magnetic superconducting heterostructures \cite{RevModPhys.77.1321,Buzdin2011,Samokhvalov2019}. It is worth noting that because of SOC, $d_x$ remains all over the N region, albeit with a very small amplitude. In contrast, the ESE in N has a decaying profile from the interface, suggesting that the proximity effect is largely dominated by OTE pair correlations. At the interface, the pair amplitudes are peaked and form larger values than for $B<\mu$, which then decay in an exponential fashion into the bulk of S. As the pair amplitudes decay, the small oscillations seen for $B<\mu$ become even smaller, which clearly shows their relation to the Zeeman field $B$, even though $B$ is only present in N; the presence of the OTE pair amplitudes is thus a signal of the inverse proximity effect. At this point, we would like to highlight that the large values of the pair amplitudes at the interface correspond to the same regime where we found trivial ABSs, see Fig.\,\ref{Fig2}(b,c,e,f). As a result, the presence of trivial ABSs can indeed enhance the OTE pair amplitudes in the helical regime $B>\mu$, even in the absence of topology or MBSs. However, exactly at zero frequency, the OTE pairing vanishes, thus revealing the topologically trivial origin of ABSs.

\begin{figure*}[!th]
    \centering
    \includegraphics[width=\textwidth]{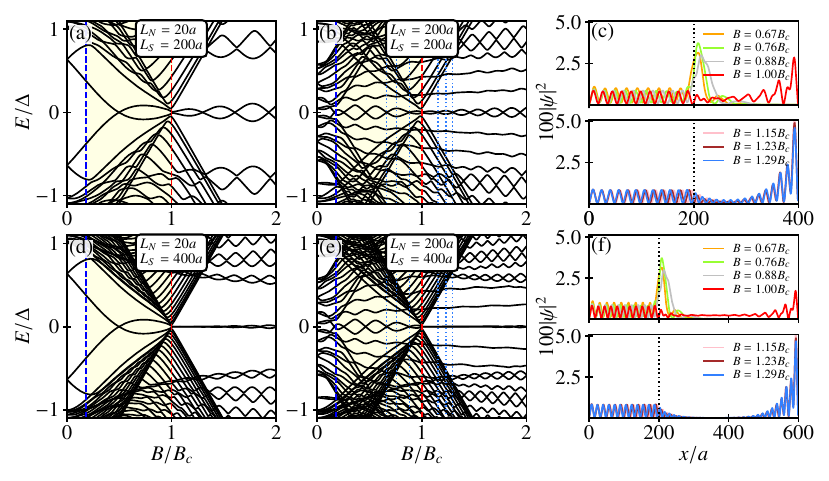}
    \caption{(a,b) Low energy spectrum for an NS junction with the Zeeman field  $B$ in N and S as a function of  $B$ for $L_{\rm N}=20a$ and $L_{\rm N}=200a$, both at $L_{\rm S}=200a$. (c) Wavefunction probability density of the state closest to zero energy as a function of the spatial coordinate at distinct values of the Zeeman field; curves in the top panel correspond to $\mu_{\rm N}<B\leq B_{\rm c}$, whereas those in the bottom panel correspond to $B>B_{\rm c}$, with $B_{\rm c}=\sqrt{\mu_{\rm S}^{2}+\Delta^{2}}$.  (d,e), and (f) Same as in (a,b) and (c), respectively,  but for $L_{\rm S}=400a$. The blue and red curves in (a,b,d,e) mark $B=\mu_{\rm N}$ and $B=B_{\rm c}$, respectively, with $\mu_{\rm N}<B<B_{\rm c}$ in yellow color;     the dotted curves in (b,e) indicate the values of $B$ at which the wavefunction probability densities are plotted in (c,f). 
Parameters:  $a=10$\,nm, $\mu_{\rm N}= 0.1$\,meV, $\mu_{\rm S}= 0.5$\,meV,  $\Delta = 0.25$\,meV, $t=25$\,meV, $\alpha = 20$\, meV nm.}
    \label{Fig4}
\end{figure*}

\section{Hybrid system with trivial zero-energy ABSs and  MBSs}
\label{section4}
We now focus on the NS junction that has a Zeeman field in both the N and S regions, see Fig.\,\ref{fig:junction}(b). In this situation, there is a topological phase transition at $B=B_{\rm c}$ that drives the S region into a topological phase for $B>B_{\rm c}$ \cite{Lutchyn,Oreg}.
Moreover, we have learned in the previous section that there is another transition happening in the NS junction:  at $B=\mu_{\rm N}$,  a helical transition appears that drives the N region into a helical phase for $B>\mu_{\rm N}$ where only two Fermi points are present \cite{PhysRevB.91.024514}. In the NS junction we consider in this part, the helical and topological phases can be achieved at distinct values of the Zeeman field, and for this reason, we choose a depleted N region with lower chemical potential than in S, namely, $\mu_{\rm N}<\mu_{\rm S}$. To provide an in-depth understanding, we here first investigate the emergence of trivial ABSs and MBSs, and then analyze their impact on the superconducting correlations. 

\subsection{Low-energy spectrum}
\label{subsection4a}
As carried out in the previous section, the energy spectrum is numerically obtained by diagonalizing the BdG Hamiltonian given by Eq.\,(\ref{eq: hamiltonian}) for the NS junction depicted in Fig.\,\ref{fig:junction}(b) and using  realistic parameters. In Fig.\,\ref{Fig4}(a,b,d,e) we present the low-energy spectrum as a function of the Zeeman field for an NS junction having short  and long N  regions as well as for two distinct values of the superconducting lengths. More precisely, we consider  $L_{\rm N}=20a$ and $L_{\rm N}=200a$ with $L_{\rm S}=200a$ and $L_{\rm S}=400a$. Moreover, Fig.\,\ref{Fig4}(c,f) shows the wavefunction probability density associated with the lowest energy state as a function of space for distinct Zeeman fields indicated by  dotted vertical lines in Fig.\,\ref{Fig4}(b,e). The first observation we highlight is that, as the Zeeman field takes finite values and increases, the NS junction undergoes two important transitions: the N region becomes helical  at $B=\mu_{\rm N}$, while the S region becomes topological   at $B=B_{\rm c}$.

When the N region is short and the S region not very long, as in Fig.\,\ref{Fig4}(a), the low-energy spectrum reveals the formation of trivial zero-energy ABSs within the helical regime (yellow shaded region) and MBSs  after the topological phase transition. In this case, the trivial (or helical) ABSs appear in the form of level crossings around zero energy, which for Fig.\,\ref{Fig4}(a) happens to emerge only a single one due to  the N region being not large. Moreover, these helical ABSs smoothly evolve towards MBSs after the topological phase transition at $B=B_{\rm c}$, which result into zero-energy oscillations due to the  superconducting length $L_{\rm S}$ being of the order of or less than twice the Majorana localization  length $\ell_{\rm M}$, namely, $L_{\rm S}\leq 2\ell_{\rm M}$.  By increasing the length of N, more energy levels appear within the superconducting gap in both the helical and topological phases. Another consequence of larger N is that more zero-energy crossings appear in the helical phase which, surprisingly, oscillate around zero energy with a decreasing amplitude as $B$ increases and approaches the topological phase transition. These helical ABSs emerge just after the helical transition  $B=\mu_{\rm N}$, which involves a kind of gap closing and reopening but still well below the topological phase $B<B_{\rm c}$; this helical transition is the same effect discussed   in Sec.\,\ref{section3}.  In the topological phase, MBSs remain largely unaffected by increasing $L_{\rm N}$ but with a larger number of  zero-energy oscillations whose amplitude increases as the Zeeman field takes stronger values. Moreover,  MBSs and helical ABSs are separated from   the first excited energy level by an energy gap of roughly the same size. The gap closing and reopening feature at $B=\mu$ as well as    the oscillatory behavior and   energy gap separation of helical ABSs can be easily confused with those of the topological phase and MBSs. However, by increasing the length of the superconducting regions S,  we find that the  helical ABSs do not change while the amplitude of the Majorana oscillations reduces, see Fig.\,\ref{Fig4}(d,e). This effect is a direct indicator that MBSs are indeed spatially nonlocal while helical ABSs are   local in space.

\begin{figure*}[!th]
    \centering
    \includegraphics[width=\textwidth]{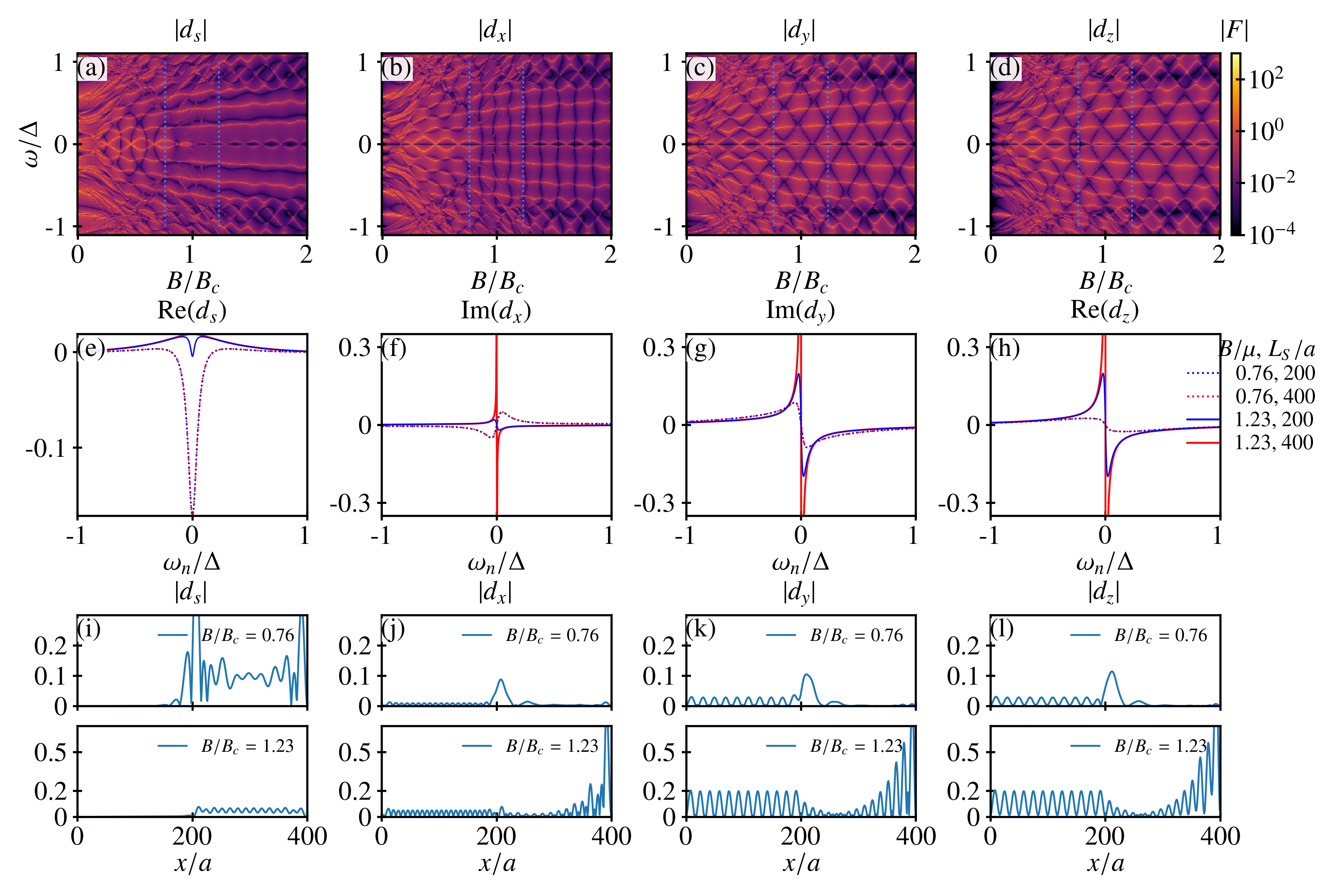}
    \caption{(a-d) Absolute value of the retarded  pair amplitudes  for an NS junction with the Zeeman field B in N and S as a function of frequency $\omega$ and  $B$ near the junction interface in N. (e-h) Pair amplitudes from (a-d) as a function of Matsubara frequency $\omega_{n}$ at $x=192a$ for two distinct values of $B$ and $L_{\rm S}$; the curves below (above) the topological transition are  depicted in dashed (solid), while for a short (long) S region the curves are shown in blue (red). (i-l) Spatial profile of the pair amplitudes in (a-d) at $\omega\approx0$ and for   $B$ below and above the topological transition. The Zeeman field $B$ in the top panel of (i-l) is fixed at $B=0.76B_{\rm c}$ which is below the helical transition, while the Zeeman field in the bottom panel is chosen above the helical transition $B=1.23B_{\rm c}$. Parameters: $x=192a$, $L_{\rm N}=200a$, while $L_{\rm S}=200a$ unless otherwise specified. The rest of the parameters are the same as in Fig.\,\ref{Fig4}.}
    \label{Fig5}
\end{figure*}

Further insights into the space dependence of the helical ABSs and MBS can be obtained from Fig.\,\ref{Fig4}(c,f) where the wavefunction probability density $|\Psi|^{2}$ for the lowest energy state is shown for distinct Zeeman fields in the helical and topological phases. In the helical phase, the wavefunction develops homogeneous oscillations of the same amplitude all over N, with a larger value at the interface, and a decaying tail towards the bulk of S. The large value of $|\Psi|^{2}$ at the interface confirms that the helical ABS is local in space and only located around one region of the NS junction. As the Zeeman field transitions into the topological phase, the wavefunction  behavior remains in the N region but is suppressed at the interface because the MBS appearing at such an interface leaks into N. Since MBSs form at both ends of S, the wavefunction probability exhibits a strong peak at the rightmost edge because that side is not coupled to anything else in contrast to the left side of S; the strong peaked profile of $|\Psi|^{2}$ already appears at $B=B_{\rm c}$. Longer S regions reduce the overlap between the Majorana wavefunctions. Nevertheless, MBS at both edges of the S region are linked in space which is at the core of their spatial nonlocality \cite{PhysRevB.86.180503,PhysRevB.86.220506,PhysRevB.96.205425,cayao2018andreev,Kuiri_2023,PhysRevB.110.224510}, unlike the behavior of the helical ABSs which are located only at the interface and hence local.

\subsection{Emergent superconducting correlations}
\label{subsection4b}
For the pair amplitudes, we obtain them following the approach presented in Subsection \ref{subsection2a} and taking into account that here the  Zeeman field is applied to both the N and S regions of the NS junction. Fig.\,\ref{Fig5}(a-d) shows the absolute value of the spin-singlet and spin-triplet retarded pair amplitudes $d_{s,x,y,z}$ as a function of frequency $\omega$ and Zeeman field at  $j=j'=192$ near the NS interface, where $L_{\rm S}=200a$ and $L_{\rm N}=200a$. Fig.\,\ref{Fig5}(e-h) depicts the pair amplitudes as functions of Matsubara frequency $\omega_{n}$ for $B$ in the helical and topological regimes and for  $L_{\rm S}\leq2\ell_{\rm M}$ and $L_{\rm S}\gg 2\ell$. Here, we remind that $\ell_{\rm M}$ is the Majorana localization length. Moreover, in Fig.\,\ref{Fig5}(i-l) we also present the space dependence of the pair amplitudes in Fig.\,\ref{Fig5}(a-d) at very low frequency $\omega\approx0$ and for $B$ in the helical and topological phases. 

A generic feature in Fig.\,\ref{Fig5}(a-d) is that the spin-singlet and all the spin-triplet pair amplitudes at the NS interface capture the formation of the low-energy spectrum, which includes the trivial ABSs in the helical phase and also MBSs in the topological phase. The spin-singlet pair amplitude has ESE symmetry, while the spin-triplet amplitudes have OTE symmetry, in both cases consistent with the Fermi-Dirac statistics of superconducting correlations, see Subsection \ref{subsection2a}. 
While all the pair amplitudes develop finite values, their behavior exhibits some differences depending on the value of the Zeeman field. The ESE pairing undergoes a reduction in intensity  at the topological phase transition $B=B_{\rm c}$, which continues in the topological phase; despite the small values, the soft intensity of the ESE pairing above $B_{\rm c}$ is able to signal the in-gap states but its roughly homogeneous intensity only weakly shows the oscillatory behavior of MBSs.    In relation to the OTE components, they acquire sizable amplitudes as $B$ takes finite values, making them coexist with the ESE pair correlations in the helical and topological phases. The coexistence of the ESE and OTE pair correlations is due to the combined effect of Zeeman field, SOC, and spatial translational invariance breaking at the NS interface \cite{odd3b,Eschrig2007,tanaka2012symmetry,cayao2019odd}.  At very low frequencies, the OTE pair amplitudes become finite   around zero frequency after the helical transition and  follow the evolution of the helical ABSs. The $d_{x}$ OTE pairing is larger in the helical phase than in the topological regime and develops vanishing values along vertical regions at almost constant Zeeman fields. In the topological phase, the OTE pair amplitudes become particularly large around zero frequency at the values of the Majorana oscillations. At exactly zero frequency, however, the OTE pair amplitudes appear to acquire vanishing values, irrespective of the trivial and topological phases. This is a surprising effect as MBSs are believed to induce an odd-frequency pairing that diverges at zero frequency \cite{Spectralbulk,cayao2019odd}.

\begin{figure}[!t]
    \centering
    \includegraphics[width=1\columnwidth]{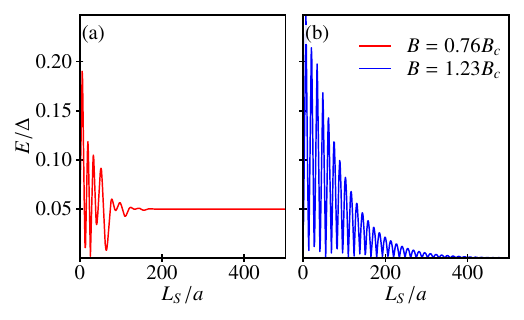}
    \caption{Energy of the lowest positive state as a function of the length of the S region $L_{\rm S}$ for distinct values of the Zeeman field $B$ in the helical ($B=0.76B_c$) and topological ($B=1.23B_c$) phases. Parameters: $L_{\rm N}=200a$. The rest of the parameters are the same as in Fig.\,\ref{Fig4}.}
        \label{Fig6}
\end{figure}

To further understand the frequency dependence of the OTE pair correlations, in Fig.\,\ref{Fig5}(e-h) we show the ESE and OTE pair amplitudes as a function of the Matsubara frequency $\omega_{n}$ for distinct lengths of S and $B$ in the helical and topological phases. We first highlight that the ESE (OTE) pair amplitude is an even (odd) function of frequency $\omega_{n}$, disregarding the value of the Zeeman field and length of S, see  Fig.\,\ref{Fig5}(e-h). While the evenness and oddness are fundamental to identify the frequency symmetry of the pair amplitudes, they strongly depend on $B$ and on the length of S. In the helical phase $B_{h}<B<B_{\rm c}$, the ESE and the OTE pair amplitudes remain unchanged when the length of the superconductor changes, see dashed blue and red curves in Fig.\,\ref{Fig5}(e-h). Here, while the ESE pairing is enhanced around zero frequency symmetrically around $\omega_{n}$, the OTE pairing disperses linearly with $\omega_{n}$ and hence vanishes at $\omega_{n}=0$, albeit it possesses large values around zero frequency. This behavior is similar to what we obtained for the OTE pairing in Fig.\,\ref{Fig3}(f-h) for the system with $B$ only in N, although therein the slope of the linear dependence is larger than the obtained here. The fact that the OTE pair amplitude in the helical regime is independent of the length of S is tied to the nature of trivial ABSs whose energy is independent of the length of S, see Fig.\,\ref{Fig6}(a). In the topological phase $B>B_{\rm c}$, the pair amplitudes strongly depend on $L_{\rm S}$, with the OTE pair amplitudes acquiring a linear dependence on $\omega_{n}$ for $L_{\rm S}=200a$ which, interestingly, then becomes divergent near $\omega_{n}\approx0$ when the superconducting region becomes longer, see Fig.\,\ref{Fig5}(e-h) for $L_{\rm S}=400a$.  This intriguing result can be understood from the fact that for $L_{\rm S}=200a$,  MBSs exhibit   oscillations around zero energy and therefore are not properly zero modes [Fig.\,\ref{Fig4}(b) and Fig.\,\ref{Fig6}(b)]; in this case their wavefunctions exhibit a spatial overlap, so MBSs are not fully nonlocal  [Fig.\,\ref{Fig4}(c)]. In contrast, for  $L_{\rm S}=400a$, MBSs become truly zero modes with their wavefunctions having a vanishing spatial overlap in space, see Fig.\,\ref{Fig4}(e,f) and also Fig.\,\ref{Fig6}(b). As a result only in the second case, MBSs are self-conjugate ($\gamma^{\dagger}=\gamma$) and therefore their normal and anomalous Green's functions are the same, and proportional to $\sim1/\omega_{n}$ \cite{PhysRevB.92.121404,cayao2019odd,PhysRevB.101.094506,PhysRevB.110.125408,PhysRevB.87.104513}, where $\gamma$ is a Majorana operator. 
 This can be seen by writing down the normal ($g$) and anomalous ($f$) Green's function associated with  the Majorana operator of a truly zero-energy MBS occurring  in very long S regions: $\gamma=\gamma^{\dagger}$ as $g(\omega_{n})=\langle \gamma^\dagger \gamma \rangle=\langle \gamma \gamma \rangle = f(\omega_{n})=1/i\omega_{n}$, where in the second equality, we have used the self-conjugation of the Majorana operator. Hence, the divergent odd-frequency  profile reveals the self-conjugation of truly MBSs. Thus, we can summarize that the OTE pairing in the topological phase can be written in terms of the Majorana energy as,
\begin{equation}
F_{\rm MBS}^{\rm O}(B>B_{\rm c})\approx \frac{i\omega_{n}}{\omega_{n}^{2}+E_{\rm MBS}^{2}}
\end{equation}
where
\begin{equation}
E_{\rm MBS}=
\begin{cases}
{\rm finite}\,,&L_{\rm S}\leq 2\ell_{\rm M}\,,\\
0\,,&L_{\rm S}\gg 2\ell_{\rm M}\,,
\end{cases}
\end{equation}
with $E_{\rm MBS}$ being the Majorana energy splitting due to a finite length of the S region, while $\ell_{\rm M}$ is the Majorana localization length. Therefore, we have that  the OTE pairing in the topological phase is divergent near $\omega_{n}\sim0$ only in the presence of truly zero-energy MBSs, which occurs in very long superconductors when the length of the superconducting region is much longer than twice the Majorana localization length. In our case, this occurs for $L_{\rm S}\geq300a$, see Fig.\,\ref{Fig6}(b).
The strong response of the OTE pairing to an increase in the length of S can also be interpreted as a signature of spatial Majorana nonlocality since MBSs appear located at the edges of the superconducting region. It is worth mentioning that this sensitivity to changes in $L_{\rm S}$ has been recently discussed in the context of zero bias conductance in similar nanowire junctions\cite{PhysRevB.108.205426}.

In terms of real space dependence of the pair amplitudes at very low frequencies, Fig.\,\ref{Fig5}(i-l) shows them at $\omega\approx0$ and at $B$ fields in the helical and topological phases. The behavior of ESE and OTE pairings in the helical phase is similar to what we obtained in Fig.\,\ref{Fig3}(i-l) when the Zeeman field is only in N, see top panels in Fig.\,\ref{Fig3}(i-l). While the OTE pairings exhibit homogeneous oscillations in N, the ESE pairing decays into the bulk of N; the finite values of superconducting correlations in N mean that all pair amplitudes contribute  to the proximity effect. At the interface, all the pair correlations are enhanced, with the OTE pairing developing  larger values that then decay in an exponentially oscillatory fashion into the bulk of S.  The finite values of OTE pairing in S are understood as the inverse proximity effect, a result of the interplay between SOC, Zeeman field, and conventional spin-singlet $s$-wave superconductivity. In the topological phase, the ESE pairing has an overall smaller profile  than the OTE pair correlations, see bottom panels in Fig.\,\ref{Fig3}(i-l).  In S, the OTE pair amplitude is particularly peaked at the edges where MBSs appear but exhibits large values only at the rightmost edge, while its amplitude is smaller at the interface due to the Majorana leakage into N. As a result, the proximity (inverse proximity) effect in the topological phase, characterized by superconducting correlations in N (S), is dominated by OTE pair correlations. This situation is completely different from what we obtained in Fig.\,\ref{Fig3}(i-l), where the OTE pair correlations are vanishingly small deep in the superconducting region.

\begin{figure}[!t]
    \centering
    \includegraphics[width=1\columnwidth]{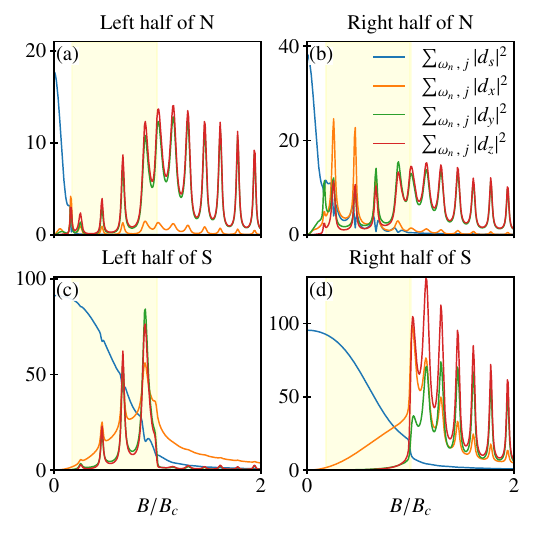}
    \caption{Squared magnitude of the spin-singlet and spin-triplet pair amplitudes $\sum_{\omega_n\,,\,j}|d_\beta(j)|^2$, with $\beta=s,x,y,z$, as 
     as a function of the Zeeman field $B$ summed over (a) the left half of  N, (b) the right half of  N, (c) the left half of S, and (d) the right half of S.  Parameters: $L_{\rm N}=200a$, $L_{\rm S}=200a$. The rest of the parameters are the same as in Fig.\,\ref{Fig4}.}
    \label{Fig7}
\end{figure}

\subsection{Zeeman dependence of the accumulated pair amplitudes}
To further understand the behavior of the emergent pair amplitudes, here we analyze their size summed over Matsubara frequencies and also over half the space in N (S), namely, $\sum_{\omega_n\,,\,j}|d_\beta(j)|^2$, with $\beta=s,x,y,z$ and $j$ denoting the lattice index. We refer to these quantities  as \emph{accumulated pair amplitudes} and we present them in Fig.\ref{Fig7} as a function of the Zeeman field $B$ for parameters where MBSs and helical ABSs are formed. The first feature to notice is that only spin-singlet pairing exists  at zero Zeeman field $B=0$. As $B$ takes nonzero values, the spin-triplet parts emerge and coexist with the spin-singlet component, developing interesting features that depend on the region where they are summed. In the left half of N,  the spin-singlet component decays rapidly as $B$ increases, reaching vanishing values at the helical phase transition when N becomes helical $B=\mu_{\rm N}$, see  Fig.\,\ref{Fig7}(a). At the same time, the spin-triplet components get finite and become even larger than their spin-singlet counterpart after the helical transition. As $B$ further increases, the spin-triplet parts oscillate with a periodicity marked by the zero energy crossings developed by the helical ABSs and MBSs. At the zero-energy crossings, the spin-triplet accumulated pair amplitudes develop large values. Interestingly, none of the spin-triplet terms exhibit a strong signature across the topological phase transition $B=B_{c}$, albeit the  $d_{y,z}$ spin-triplet parts become visibly dominant [Fig.\,\ref{Fig7}(a)]. This behavior is preserved when the accumulated pair amplitudes are obtained in the right half of N, shown in Fig.\,\ref{Fig7}(b), with the difference that the spin-singlet part has finite values over the helical phase but still smaller in the topological regime. 

For the accumulated pair amplitudes in the S regions, we find that the spin-singlet part decays slower with $B$ than when summed in N, having finite values in the helical regime and  smaller values after the topological phase transition, see Fig.\,\ref{Fig7}(c,d). When it comes to the spin-triplet components, when summed over the left half they exhibit large values in the helical phase at the energies of the helical ABSs, while smaller in the topological phase that signals a visible feature of the topological phase transition [Fig.\,\ref{Fig7}(c)]; in both cases, the oscillations are determined by the helical ABSs and MBSs. When summed over the right half of S, the accumulated spin-triplet pair amplitudes do not exhibit  oscillations in the helical regime but, interestingly, only in the topological phase and due to MBSs [Fig.\,\ref{Fig7}(d)]. These oscillations in the spin-triplet components in the topological phase occur after a huge enhancement   at the topological phase transition $B=B_{\rm c}$ and a sudden drop of the spin-singlet part. As a result, the accumulated spin-triplet pair amplitudes summed over the right half of S signal the topological phase transition and the formation of MBSs, while summed over N  they unveil the helical phase transition and helical ABSs.  

\subsection{Signatures of odd-frequency pairing in experiments}
\label{subsection4c}
Direct measurement of odd-frequency pairing is a challenging task because it implies measuring the pair amplitude $F$, see Refs.\,\cite{tanaka2012symmetry,RevModPhys.77.1321,RevModPhys.91.045005,cayao2019odd,triola2020role}. However, there exist   several experimental signatures that can indirectly reveal its presence in superconductor-semiconductor hybrid systems, as we discuss below.

One way to measure the OTE pairing discussed here is by using Andreev spectroscopy, which detects the Andreev reflection probability. Since the squared magnitude of the anomalous Green's function, $|F|^2$, is directly related to Andreev reflection probability, OTE pairing can be accessed through Andreev spectroscopy. Note that our results suggest that Andreev spectroscopy would be most effective when performed at the rightmost side of the superconductor in the topological phase, where the OTE pairing signal is strongest due to the localization of Majorana bound states.  Furthermore, the behavior of spin-triplet pairing in relation to external Zeeman fields can help distinguish between trivial ABSs and topological MBSs. As shown in Fig.\,\ref{Fig7}, the accumulated spin-triplet pair amplitudes exhibit distinct signatures at the topological phase transition that are particularly pronounced at the edges of the superconducting region. However, we note that, under generic circumstances, both even- and odd-frequency pairings contribute to the Andreev transport \cite{PhysRevB.109.L100505}, making it difficult to isolate the OTE pairing signal. Nevertheless, there already exist ways to identify the contribution from odd-frequency pairing  via Andreev transport,  e.g. in multiterminal junctions \cite{PhysRevB.109.205406,PhysRevB.93.201402}. Thus, albeit challenging, the OTE pairing can be detected by means of Andreev spectroscopy in superconductor-semiconductor hybrid systems.

Besides Andreev spectroscopy, the OTE pairing we discuss can also be measured by means of the anomalous proximity effect  in diffusive normal-superconductor junctions \cite{PhysRevB.87.104513,D2NR05864B,PhysRevB.72.140503,PhysRevB.94.054512,Proximityp,PhysRevB.71.094513,PhysRevLett.99.067005,PhysRevB.102.140505}. In conventional junctions, the local density of states (LDOS) in the normal region is suppressed at energies below the superconducting gap. However, when OTE correlations dominate since they are robust against disorder, they induce an enhancement in the zero-energy LDOS in the normal region \cite{tanaka2012symmetry}, which can be detected through tunneling spectroscopy. Thus, to detect the OTE discussed in the previous subsections by means of the anomalous proximity effect, the N region of the considered geometries needs to be appropriately disordered.

Yet another approach is offered by the Josephson effect  \cite{odd3,Yokoyama2008} since it involves the transfer of Cooper pairs between superconductors which are determined by  emergent pair amplitudes. A scanning tunneling microscope experiment using a superconducting tip can probe the current-phase relation of a Josephson junction, which differs significantly when odd-frequency correlations are present. This is because the Josephson current  is sensitive to the pairing symmetry and can reveal the presence of OTE components. 
To implement this detection approach, it would require  extending the considered geometries into Josephson junctions with a finite phase difference between superconductors. Due to symmetry constraints, Josephson current will flow when the left and right superconductors host the same type of Cooper pairs, enabling a way to identify the OTE correlations.

\section{Effect of Zeeman field orientation on the helical and topological regimes}
\label{subsection4d}
\begin{figure*}[!t]
    \centering
    \includegraphics[width=1\textwidth]{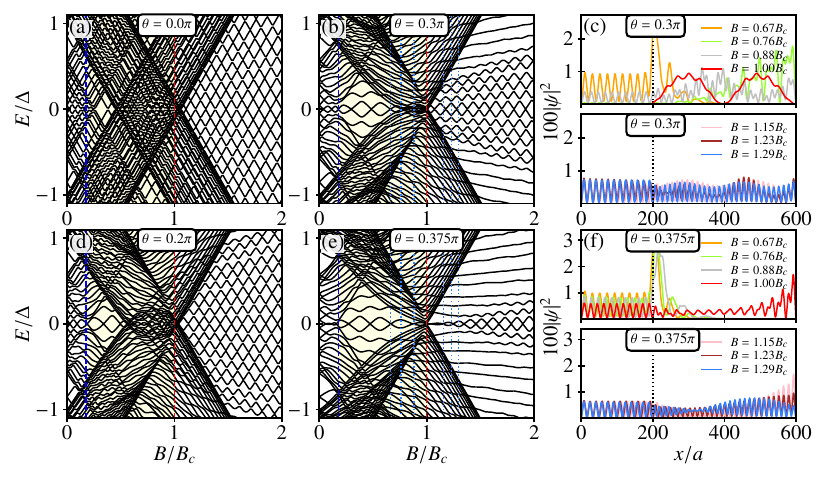}
    \caption{Low-energy spectrum for an NS junction with the Zeeman field $B$ in N and S as a function of $B$ at distinct values of the polar angle $\theta$ of the Zeeman field: (a) $\theta=0$, (b) $\theta=0.3\pi$, (c) $\theta=0.2\pi$, and (d) $\theta=0.375\pi$. Wavefunction probability densities of the state closest to zero energy as a function of the spatial coordinate at $\theta=0.3\pi$ (c) and $\theta=0.375\pi$ (f). The blue and red curves mark $B=\mu_{\rm N}$ and $B=B_{\rm c}$, respectively. The dotted curves in (b,e) indicate the values of $B$ at which the wavefunction probability densities are plotted in (c,f). Parameters: $L_{\rm N}=200a$, $L_{\rm S}=400a$, and $\phi=0$. The rest of the parameters are the same as in Fig.\,\ref{Fig4}.}
    \label{Fig a1}  
\end{figure*}
So far, we have considered a Zeeman field aligned along the $x$-direction, perpendicular to the Rashba SOC which points along the $z$-direction. Here, we discuss how rotating the Zeeman field affects both the helical and topological phases. To that end, we modify the Zeeman term in the Hamiltonian to the following:
\begin{equation}
\label{HB}
\begin{split}
    H_{\rm B} &= -B\sum_{j=1}^N\psi_j^\dagger \Big(\sin(\theta)\cos(\phi)\tau_z\sigma_x     \\
    &+\cos(\theta)\tau_z\sigma_z+\sin(\theta)\sin(\phi)\tau_0\sigma_y \Big)\psi_j\,,
    \end{split}
\end{equation}
where  $\theta$ and $\phi$ are the polar and azimuthal angles respectively.  First, we note that changing $\phi$ rotates the Zeeman field around the Rashba SO axis, which does not affect the magnitude of the parallel and orthogonal components of the Zeeman field along the Rashba SO axis. Thus, both the topological and helical phases are unaffected by the angle $\phi$.  On the other hand, when the Zeeman field is tilted at an angle $\theta$, the helical and topological phases exhibit a strong angular dependence. Of particular relevance is that there exists a critical angle $\theta_c$ at which the bulk system becomes gapless, which corresponds to the situation when the Zeeman field component parallel to the SO axis is equal to the pair potential, namely, $B_{z}=\Delta$, where $B_{z}$ is the $z$-component of the Zeeman field given in Eq.\,(\ref{HB}). Thus, the critical angle is obtained as $\theta_c=\arccos(\Delta/B)$, and is expected in the helical and topological phases. The impact of $\theta$ and its critical value is shown in Fig.\,\ref{Fig a1}, where we present the Zeeman evolution of the energy spectrum and the space dependence of the lowest states. When the Zeeman field is parallel to the SO axis ($\theta=0$), the helical phase disappears completely, as seen in Fig.\,\ref{Fig a1}(a). This occurs because the Zeeman field and SOC  cannot compete to create  a helical gap with counter-propagating states necessary for helicity \cite{cayao2024_JD}. As $\theta$ increases from 0 to $\pi/2$, the helical region extends [Fig.\,\ref{Fig a1}(b,d,e)], reaching its maximum size when the Zeeman field is perpendicular to the SO axis ($\theta=\pi/2$); see also Fig.\,\ref{Fig4}(e). For $\pi/2<\theta<\pi$, the helical region decreases until it vanishes again at $\theta=\pi$. The topological phase also exhibits a critical dependence on the orientation of the Zeeman field, mainly because it becomes  gapless at  $\theta_c$, implying the absence of   topological protection when $B>B_{c}$.   The topological phase transition, however, is not affected by $\theta$ and still occurs at $B=B_{c}$, consistent with previous studies \cite{tilting2014,tilting2014b,cayao2016hybrid,Mondal2025}.  Thus, the topological region with a finite gap in parameter space is confined to the angular range $\theta_c<\theta<\pi-\theta_c$.

\begin{figure}[!t]
    \centering
    \includegraphics[width=1\columnwidth]{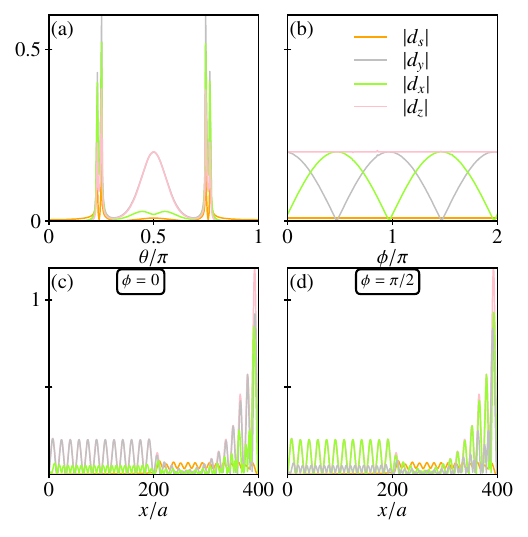}
    \caption{Absolute value of the pair amplitudes $|d_\beta|$ in the topological phase at $\omega\approx0$ near the NS interface in N ($x=192a$) as a function of $\theta$ at $\phi=0$ (a) and as a function of  $\phi$ at $\theta=0.5\pi$ (b). (c) Spatial profile of the pair amplitudes at $\omega\approx0$ and for $\theta=\pi/2$ and $\phi=0$.  (d) Same as in (c) for $\phi=\pi/2$. Parameters: $x=192a$, $B=1.23B_{\rm c}$, $L_{\rm N}=200a$, and $L_{\rm S}=200a$. The rest of the parameters are the same as in Fig.\,\ref{Fig4}.}
    \label{Fig a4}
\end{figure}

To further examine the effect of $\theta$, in Fig.\,\ref{Fig a1}(c,f) we show the spatial profile of the wavefunction probability density of the lowest energy state at different $\theta$ in the trivial and topological phases. When $\theta<\theta_c$, the wavefunction probability density in both the helical and topological phases is delocalized and is spread out across the entire junction, see Fig.\,\ref{Fig a1}(c). This is because the system is gapless for $\theta\notin [\theta_c,\pi-\theta_c]$. Interestingly, the wavefunction probability density in the topological phase has a homogeneous oscillatory profile while in the helical phase the oscillatory profile is inhomogeneous. For $\theta \in [\theta_c,\pi-\theta_c]$, when the system is gapped and shown in Fig.\,\ref{Fig a1}(f), we recover the wavefunction behavior shown in Fig.\,\ref{Fig4}(c,f) for the helical and topological phases. Specifically, for $\theta$ such that the system is gapped, the lowest energy wavefunctions in the helical and topological phases are localized at the NS interface and edges of the S region, respectively, see Fig.\,\ref{Fig a1}(f).

The orientation of the Zeeman field also induces strong effects on the emergent pair amplitudes $d_\beta$. To show these consequences, we first inspect the dependence of $|d_\beta|$ on the polar angle $\theta$ in the topological phase,  which we present in Fig.\,\ref{Fig a4}(a) at $\phi=0$ and $\omega\approx0$ near the NS interface in N. We see that, for $\theta\notin [\theta_c,\pi-\theta_c]$ when the system is gapless, both the singlet and triplet pair amplitudes are suppressed, indicating the absence of MBSs. Exactly at $\theta=\theta_c,\pi-\theta_c$,   the OTE pair amplitudes develop sharp peaks, signaling the suppression of the energy gap. As we increase the angle above $\theta_c$ and enter the region $\theta\in[\theta_c,\pi-\theta_c]$, where the system is gapped, the OTE pair amplitudes initially decrease but then increase, with $d_{z,y}$ reaching a maximum when the Zeeman field becomes orthogonal to the SO axis.  In contrast, the ESE pair amplitude $d_{s}$ acquires small values regardless of the value of $\theta$. Consequently, the OTE pair amplitudes are the dominant pair correlations in the normal region near to the NS interface. We have verified that the behavior shown in Fig.\,\ref{Fig a4}(a) remains unchanged when varying  the azimuthal angle $\phi$. This robustness makes it possible to probe OTE signatures (and by extension, the presence of Majorana or trivial ABS physics) using methods such as Andreev spectroscopy.

Since the OTE pair amplitudes are most pronounced when the Zeeman field is perpendicular to the spin-orbit coupling, in Fig.\,\ref{Fig a4}(b) we show the absolute value of the ESE and OTE pair amplitudes of Fig.\,\ref{Fig a4}(a) as a function of $\phi$ at $\theta=\pi/2$ in the topological phase. Here, $d_x$ and $d_y$ develop an  oscillatory dependence   as a function of $\phi$ with a period of $\pi$. Also, $d_y$ is shifted by $\pi/2$ with respect to $d_x$, but both $d_y$ and $d_x$ have a similar amplitude with their maxima occurring at $\phi=0$ and $\phi=\pi/2$, respectively. On the other hand, the $d_z$ pair amplitude remains constant with its value being the same as the maximum value of $d_x$ and $d_y$. The ESE pair amplitude $d_s$ is also constant and has a very small value, indicating that the singlet pair amplitude is suppressed. The spatial profiles of the pair amplitudes at $\omega\approx0$ for $\phi=0$ and $\phi=\pi/2$ are shown in Fig.\,\ref{Fig a4}(c) and Fig.\,\ref{Fig a4}(d), respectively. We see that when the Zeeman field is along the $x$ axis ($\phi=0$), $d_x$ is suppressed in N . For the Zeeman field is along the $y$ axis ($\phi=\pi/2$), $d_y$ is suppressed in N. This is a manifestation of short- and long-range spin-polarized supercurrents, discussed before in the context of FS junctions \cite{RevModPhys.77.1321,Buzdin2011,Samokhvalov2019}. We have also verified that the spatial profiles of both $d_z$ and $d_s$ are not affected by the azimuthal angle $\phi$ and remain constant; $d_s$ is suppressed in N and has a finite value in S, while $d_z$ is enhanced  at the edges of the S region in the presence of MBSs.  To close this part, we highlight that the angular dependencies discussed here provide additional means to distinguish between helical and topological phases based on the emergent superconducting correlations. 
In particular, by systematically rotating the applied Zeeman field, it might be possible to identify the behavior of OTE and ESE pair correlations, as well as their contribution to conductance, under the presence of helical ABSs and MBSs. This rotation protocol could complement the  experimental signatures discussed in Subsection\,\ref{subsection4c}.

\section{Conclusions}
\label{section5}
In conclusion, we have investigated the emergence of superconducting correlations in superconducting systems hosting Majorana and trivial zero-energy states. In particular, we have considered realistic normal-superconductor junctions with spin-orbit coupling, where  an external Zeeman field   drives the system into a helical phase with trivial zero-energy Andreev bound states and into a topological phase with Majorana bound states. We have shown that the trivial Andreev bound states are local in space appearing at interfaces, while Majorana bound states emerge located at both sides of the superconducting regions and are thus nonlocal in space. We have argued that this space profile makes Majorana bound states strongly sensitive to variations in the superconducting length, which is important for distinguishing them from the helical Andreev bound states.

We have then demonstrated that normal-superconductor junctions with spin-orbit coupling under Zeeman fields locally host even-frequency spin-singlet even-parity  and odd-frequency spin-triplet even-parity pair correlations, whose coexistence determines both the proximity and the inverse proximity effects. In the helical phase, we found that the odd-frequency spin-triplet pairing can be enhanced due to the formation of trivial zero-energy Andreev bound states and hence without any relation to topology. This odd-frequency pairing, however, turns out to be linear in frequency and, despite its large value around zero frequency, it vanishes exactly at zero frequency irrespective of the superconducting length; this behavior thus signals the topologically trivial nature of the helical phase. In the topological phase, we have demonstrated that the odd-frequency spin-triplet pairing is large in the presence of Majorana bound states, acquiring a divergent frequency dependence at around zero frequency only when Majorana bound states are truly zero modes, which is achieved in superconductors whose length is larger than twice the Majorana localization length. We have shown that in short superconductors and in the topological phase, the odd-frequency pairing, although large due to Majorana bound states, linearly vanishes at zero frequency as in the helical regime. The divergent   profile of the odd-frequency spin-triplet pairing in the topological phase under truly zero-energy Majorana bound states can therefore be seen as a strong signature of Majorana nonlocality, which is tied to the inherent self-conjugation   of Majorana bound states.  Furthermore, we have shown that, when the Zeeman field has a component parallel to the spin-orbit axis, the superconducting correlations strongly depend on the orientation angle of the Zeeman field. Specifically, we have found that the odd-frequency spin-triplet   components become most pronounced when the Zeeman  field is perpendicular to the spin-orbit   axis. Moreover, the $d_x$ and $d_y$ pair amplitudes exhibit an oscillatory behavior as a function of the azimuthal angle of the Zeeman field, reflecting the role of the Zeeman field components on the emergent correlations. Our findings  provide a comprehensive understanding of the impact of  Majorana and trivial zero-energy Andreev bound states on the emergent superconducting correlations in superconductor-semiconductor hybrids.



\section{Acknowledgements} 
We thank T. Mizushima and N. Nagaosa for insightful discussions. E. A. acknowledges financial support from Nagoya University and Mitsubishi Foundation. S. T. acknowledges the support from JSPS with Grants-in-Aid for Scientific research (KAKENHI Grants No.~24H00853).  Y. T. acknowledges support from JSPS with Grants-in-Aid for Scientific Research  (KAKENHI Grant No. 23K17668  and 24K00583). J. C. acknowledges financial support from the Japan Society for the Promotion of Science via the International Research Fellow Program,  the Swedish Research Council (Vetenskapsr{\aa}det Grant No. 2021-04121),  and the Carl Trygger’s Foundation (Grant No. 22: 2093). 

\section{Data Availability}
The data that support the findings of this article are not publicly available because the cost of preparing, depositing, and hosting the data would be prohibitive within the terms of this research project. The manuscript includes sufficient information to reproduce the data. The data are available from the authors upon reasonable request.

\bibliography{biblio}

\end{document}